# An efficient multi-use multi-secret sharing scheme based on hash function

Angsuman Das & Avishek Adhikari
Department of Pure Mathematics
University of Calcutta
35, Ballygunge Circular Road
Kolkata-700019
Email- angsumandas054@gmail.com & aamath@caluniv.ac.in

**Abstract**

*In this paper, a renewable, multi-use, multi-secret sharing scheme for general access structure based on one-way collision resistant hash function is presented in which each participant has to carry only one share. By applying collision-resistant one-way hash function, the proposed scheme is secure against conspiracy attacks even if the pseudo-secret shares are compromised. Moreover, high complexity operations like modular multiplication, exponentiation and inversion are avoided to increase its efficiency. Finally, in the proposed scheme, both the combiner and the participants can verify the correctness of the information exchanged among themselves.*

## 1  Introduction

In recent days, mathematics have been widely used in public-key cryptography, which in turn has been widely used in various applications such as e-cash, e-voting etc. One of the important issues of the public key cryptosystem is the key management and thus the private key of the public-key infrastructure should be safely preserved.

A *secret sharing scheme* (SSS) allows to split a secret $s$ into different pieces, called *shares*, which are given to the set of participants $\mathcal{P}$, such that only certain qualified subsets of participants can recover the secret using their respective shares. The collection of those qualified set of participants is called *access structure* $\Gamma_s$ corresponding to $s$. Blakley [1] and Shamir [9], in 1979, independently, came out with a scheme known as $(t, n)$ threshold secret sharing scheme. Stadler [10] proposed a verifiable secret sharing scheme for general access structure. However, the schemes [1], [9], [10] dealt with single secret and once the secret was updated with a new one, the system had to reissue new share to each participant. This may be considered as system resources consuming and sometimes impracticable. To eliminate these weaknesses, in 1994, He-Dawson [5] proposed a multistage $(t, n)$ threshold secret sharing scheme. In 2007, Geng *et al.* [4] proposed a multi-use threshold secret sharing scheme using one-way hash function and pointed out that the He-Dawson scheme was actually an one-time-use scheme and can't endure conspiring attacks. A SSS is said to be *multi-use* if even after a secret is reconstructed by some participants, the combiner cannot misuse their submitted information to reconstruct some other secrets. To make a scheme multi-use, the participants do not provide the combiner with the original share but a shadow or image of that share, which is actually an entity that depends on the original share. This image or shadow is known as the *pseudo-secret share*. In 2006, Pang *et al.* [8] proposed a multi-secret sharing scheme for general access structure in which all the secrets are revealed at a time. In 2008, Wei *et al.* [12] proposed a renewable secret sharing scheme for general access structure. The proposed scheme also allows new secrets to be added. In addition, the participant set and the access structure can be changed dynamically without updating any participant's share. A SSS is said to be *verifiable* if the participants can check the correctness of their shares given by the dealer and the reconstructed secret given by the combiner and the combiner can check whether the participants have submitted their correct pseudo-shares or not. The proposed scheme is a verifiable, multi-secret sharing scheme where each secret can be reconstructed independently and different secrets corresponding to different access structures may be shared. The uses of only 'XOR' operation and the hash function make the scheme efficient compared to the schemes [9], [10] which use modular multiplication, exponentiation and inversion.

The rest of this paper is organized as follows: The proposed scheme is discussed in section 2. The analysis and discussions on the proposed scheme are given in section 3 and finally, the conclusion is given in section 4.

## 2   The proposed scheme

In this section, we present a new efficient, renewable, multi-use multi-secret sharing scheme for general access structure using one-way collision resistant hash function [11].
**Aim of the scheme :**
Suppose, $\mathcal{P}=\{P_1, P_2, \ldots, P_n\}$ be a set of $n$ participants and $s_1, s_2, \ldots, s_k$ be the $k$ secrets to be shared by a trusted dealer $\mathcal{D}$ such that $s_i \in \{0,1\}^q$ for $i = 1, 2, \ldots, k$ with access structures $\Gamma_{s_i} = \{A_{i1}, A_{i2}, \ldots, A_{it_i}\}$ where $\{0,1\}^q$ denotes the set of all binary strings of fixed length $q$ and $A_{il}$ is the $l$th qualified subset of the access structure of $i$th secret $s_i$.

The scheme consists of three basic phases,
**Dealer's phase**
**Step 1:** The dealer $\mathcal{D}$ chooses:
*(i)* $H$, a suitable secure collision resistant one-way hash function, which takes as argument a binary string of arbitrary length and produces as output a binary string of a fixed length $q$, where $q$ is the length of each secret.
*(ii)* $x_\alpha \in_R \{0,1\}^q$, $\alpha = 1, 2, \ldots, n$, where '$\in_R$' denotes the random selection.
**Step 2:** The dealer $\mathcal{D}$ sends $x_\alpha$ to $P_\alpha$ secretly, for $\alpha = 1, 2, \ldots, n$ and publishes $H$ and the access structures $\Gamma_{s_i}$, for $i = 1, 2, \ldots, k$. The selection of $x_\alpha$ in *(ii)* of *Step 1* may also be done by the participants and they themselves may send their shares to the dealer through a secure channel.
**Pseudo-share generation phase**
Let $l = [log_2 k] + 1$ and $m = [log_2 t] + 1$, where $t = max\{t_1, t_2, \ldots, t_k\}$ and $t_i = |\Gamma_{s_i}|$, as explained in **Aim of the scheme**.
**Step 1:** For $i = 1, 2, \ldots, k; j = 1, 2, \ldots, t_i$; the dealer computes
$$S_{ij} = s_i \bigoplus \{\bigoplus_{\alpha: P_\alpha \in A_{ij}} H(x_\alpha||i_l||j_m) \}$$
where $i_l$ denotes the $l$-bit binary representation of $i$, $j_m$ denotes the $m$-bit binary representation of $j$, '||' denotes the concatenation of two binary strings and $\bigoplus$ denotes the XOR

operation, i.e., componentwise addition modulo 2.

**Step 2:** $\mathcal{D}$ publishes the values $S_{ij}, H(s_i), H^2(x_\alpha||i_l||j_m)$ for $\alpha = 1, 2, \ldots, n$; $i = 1, 2, \ldots, k$; $j = 1, 2, \ldots, t_i$, where $H^2(x_\alpha||i_l||j_m)$ means $H(H(x_\alpha||i_l||j_m))$.

**Combiner's phase**

Suppose the group of participants $A_{ij}$ of $\Gamma_{s_i}$ submit their shares to the combiner to get $s_i$. Then the combiner can check whether a particular participant has given his pseudo-secret share $H(x_\alpha||i_l||j_m)$ correct or not, by verifying it with the corresponding public value $H^2(x_\alpha||i_l||j_m)$.

If each of the pseudo-secret shares is correct, the combiner $\mathcal{C}$ calculates
$$S_{ij} \bigoplus \{\bigoplus_{\alpha: P_\alpha \in A_{ij}} H(x_\alpha||i_l||j_m)\}$$
which is eventually equal to $s_i$.

The participants in $A_{ij}$ of $\Gamma_{s_i}$ can check whether the combiner is giving them back the correct secret $s_i$ or not, by verifying it with the public value $H(s_i)$.

## 3 Analysis of the scheme

### 3.1 Security of the scheme

We discuss the security of the scheme with respect to the pseudo-secret shares, the shares and the secrets.

**1. Security of the pseudo-secret shares:** An adversary $\mathcal{A}$ can try to derive participant's pseudo-secret share from $H^2(x_\alpha||i_l||j_m)$, which is public. But if $\mathcal{A}$ succeeds in doing that, then $\mathcal{A}$ will be able to find a pre-image of an element under $H$, which is assumed to be computationally hard.

**2. Security of the shares:** An adversary $\mathcal{A}$ can try to derive participant's share from a previously submitted pseudo-secret share $H(x_\alpha||i_l||j_m)$. But as the shares are chosen by the dealer randomly and passed on to the respective participants secretly, adversary $\mathcal{A}$ would have to invert the hash function $H$, which is assumed to be computationally hard.

**3. Security of the secrets:** Suppose all, but one participant, in $A_{ij}$ comes to get $s_i$. They have to guess the pseudo-secret share of the missing participant from $\{0,1\}^q$, where $q$ is the fixed bit-length of the hashed value. So, they have $2^q$ choices. Whereas, a layman, without any share, who knows only that the secret $s_i$ is a $q$-bit string, has also $2^q$ many choices for the secret. Thus, a forbidden set of participants has no extra privilege than an outsider. Same thing happens if any other unauthorised subset of participants comes to reconstruct any secret. Thus, the scheme is computationally secure under the security of the chosen collision resistant hash function $H$.

**Remark :** Note that, in the proposed scheme, the size of the secret space is same as that of the share space.

### 3.2 The scheme is a multi-use one

Suppose, a participant $P_\alpha$ submits his pseudo-secret share to the combiner for the reconstruction of a particular secret $s_i$. Again, let the same participant $P_\alpha$ be present in the access structure $\Gamma_{s_j}$, $i \neq j$. If his pseudo-secret share in both cases are same, then the combiner may misuse his share without his consent while reconstructing $s_j$. Thus, the pseudo-secret shares of a participant for different secrets and even for different qualified subsets for the same secret should be different.

Various schemes e.g., [5], [4] and [7], incorporate succesive use of hash functions to deal with this situation, which contribute to greater complexities of the schemes. In the present scheme, that bottleneck is removed by the use of concatenation of binary strings of $i$ and $j$ to $x_\alpha$ and as a result, it is sufficient to use the hash function only once.

## 3.3 Renewal of the scheme

In a practical scenario, it may be necessary to add new secrets and corresponding access structure. In addition, it may be required to change the participant set or the access structure corresponding to some secret(s).

In the proposed scheme, these changes can be done dynamically without updating any participant's share. This can be achieved by the dealer by simply modifying the pubic values $S'_{ij}$ and $\Gamma_{s'_i} = \{A'_{i1}, A'_{i2}, \ldots, A'_{it_i}\}$, where $s'_i$ is the added secret and/or $\Gamma_{s'_i}$ is the modified access structure.

## 3.4 Performance analysis

The present scheme is an efficient one due to the following reasons:
**(1)** The only operations that are used are XOR and finding the hashed values, out of which the former is of negligible complexity. As hash function plays an important role in the proposed scheme, we calculate the number of times for which the hash function, $H$ is used by each of the participants, the dealer and the combiner for a single secret $s_i$ in the worst possible case.

By Dealer: $nt$ times for calculating $S_{ij}$, $nt$ times for calculating $H^2(x_\alpha||i_l||j_m)$ and once for publishing $H(s_i)$, thereby totaling to $2nt + 1$ times.

By each participant: Once for calculating $H(x_\alpha||i_l||j_m)$ and once for verifying $H(s_i)$.

By Combiner: $n$ times for checking the correctness of the submitted pseudo-secret shares by calculating $H^2(x_\alpha||i_l||j_m)$.

**(2)** In [5], [4] and [7], successive use of hash function is incorporated to use the same share more than once, but in the present paper, the same issue is resolved just by using the hash function only once.

**(3)** In the present scheme, modular exponentiation, modular multiplication and modular inversion are not used anywhere, contrary to various schemes where some of them e.g., [10], [12], used modular exponentiation and while others e.g., [9], [8], [12] used lagrange's interpolation techniques for the reconstruction of secrets. So, the computational cost in the present scheme is quite low compared to other schemes using the above-mentioned operations.

The schemes [8] and [12] are also having almost the same features as that of the proposed scheme. We give a brief comparison of these two schemes with the proposed one in table(1)

## 3.5 Comments on dealer verification

In reality, the dealer should also be verifiable as due to dishonesty of the dealer, some/all of the participants may be deprived from reconstructing the original secret or secrets. To deal with this crisis, there are many schemes [10], [6], [2], [8], in present literature which allows dealer verification.

In most of the existing schemes allowing dealer verification, the issue of verification have been dealt with using exponentiation (say, to the base $g$, where $g$ is a primitive element of the underlying group) e.g., [10], [6], [2] or by applying one-way hash function (say $H$) e.g., [8] on the secret shares. Let us have a look on the matter from a broader platform: suppose $x$ is the share of a participant given by the dealer. In the above stated schemes, the dealer either publishes $H(x)$ or $g^x$ to enable the participants to verify their own shares. Now, three cases may arise.

Table 1: Comparison among [8], [12] and the proposed scheme w.r.t various parameters

| Features | Proposed Scheme | Pang et.al [8] | Wei et.al [12] |
|---|---|---|---|
| Multi-secret | Yes | Yes | Yes |
| Access Structure | General | General | General |
| Secret revealing order | Any | Predetermined (All at a time) | Predetermined (Fixed order) |
| Use of interpolation | No | Yes | Yes |
| Use of modular exponentiation | No | No | Yes |
| Use of hash/ one-way function | Yes (hash function) | Yes (hash function) | No (DLP is used) |

***Case 1:*** Suppose, instead of $x$, the dealer calculates relevant quantities and constructs the scheme using $x'$, and sends $x$ ($x \neq x'$) as the share and publishes $H(x)$ (or $g^x$). Then, though the participant will be ensured that his share is valid by checking the verification procedure, the qualified set where he belongs will not get the correct secret back. Thus, the dealer can forge a participant without being noticed. So, in reality, the dealer is not verified using these methods.

***Case 2:*** In another scenario e.g., [3], where the participants choose their share $x$ themselves and send it to the dealer, the same problem may arise.

***Case 3:*** In the worst case, it may happen that the dealer publishes the wrong entities like, the hash function $H$ or the prime $p$, which are involved in initial set up of the scheme. In that case, the total system will collapse. How can these quantities be checked to be correct or not?

So, in our scheme, we assume the dealer $\mathcal{D}$ to be a trusted one. But, it can be a noble issue to search for techniques which resolve the aforesaid problems and we invite researchers to have a look on it.

## 4 Conclusion

In this paper, we have presented a multi-secret sharing scheme with general access structures based on one-way collision resistant hash function. The major characteristics of its construction are multi-use of the shares and that different secrets can be reconstructed according to their access structure, which provide more flexibility. It has been emphasised that, unlike several other authors, operations like modular multiplication, exponentiation and inversion are not used, thereby reducing the computational cost of the scheme to quite a large extent. By applying one-way hash function and the concatenation operation, the proposed scheme is secure against notorious conspiracy attacks even if the pseudo-secret shares are compromised. Analysis showed that this proposed scheme is an efficient one and it can provide great capabilities for many applications, such as e-voting, multi-party protocols, oblivious transfer, privacy preserving data-mining etc.

## 5 Acknowledgement

The first author is supported by National Board of Higher Mathematics, Department of Atomic Energy, Government of India (No- 39/1/2008-RD-II/3881).